\documentclass{sig-alternate-per}

\usepackage[nolist]{acronym}                    
\usepackage{url}
\usepackage{breakurl}

\usepackage{listings}
\usepackage{enumitem}
\usepackage[font=scriptsize]{subfig}

\begin{document}
%

\title{Specification of Complex Structures in Distributed \\ Service Function 
Chaining Using a YANG Data Model}
%
%
%
%
%

\numberofauthors{2} 
%
\author{
%
%
\alignauthor
Sevil Mehraghdam\\
       \affaddr{University of Paderborn}\\
       \affaddr{33098 Paderborn, Germany}\\
       \email{s.mehraghdam@upb.de}
\and
\alignauthor
Holger Karl\\
       \affaddr{University of Paderborn}\\
       \affaddr{33098 Paderborn, Germany}\\
       \email{holger.karl@upb.de}
}

\begin{acronym}[WWWW]

\acro{3gpp}[3GPP]{3rd Generation Partnership Project}

\acro{acl}[ACL]{Access Control List}
\acro{adc}[ADC]{Application Delivery Controller}
\acro{adt}[ADT]{Application Deployment Toolkit}
\acro{alg}[ALG]{Application-Level Gateway}
\acro{av}[AV]{Anti-virus}

\acro{bnf}[BNF]{Backus-Naur Form}
\acro{bng}[BNG]{Broadband Network Gateway}


\acro{cache}[CACHE]{Cache}
\acro{cgnat}[CGNAT]{Carrier-Grade Network Address Translation}
\acro{cr}[CR]{Core Router}
\acro{crm}[CRM]{Customer Relationship Management}

\acro{dpi}[DPI]{Deep Packet Inspector}

\acro{ebnf}[EBNF]{Extended Backus-Naur Form}
\acro{erp}[ERP]{Enterprise Resource Planning}
\acro{etsi}[ETSI]{European Telecommunications Standards Institute}
\acro{epc}[EPC]{Evolved Packet Core}

\acro{fpga}[FPGA]{Field Programmable Gate Array}
\acro{fw}[FW]{Firewall}

\acro{ggsn}[GGSN]{Gateway GPRS Support Node}

\acro{http}[HTTP]{Hypertext Transfer Protocol }

\acro{ids}[IDS]{Intrusion Detection System}
\acro{ietf}[IETF]{Internet Engineering Task Force}
\acro{ip}[IP]{Internet Protocol}
\acro{isg}[ISG]{Industry Specification Group}



\acro{lb}[LB]{Load Balancer}
\acro{li}[LI]{Lawful Interception}


\acro{miqcp}[MIQCP]{Mixed Integer Quadratically Constrained Program}

\acro{nat}[NAT]{Network Address Translator}
\acro{nf}[NF]{Network Function}
\acro{nfv}[NFV]{Network Function Virtualization}


\acro{pcef}[PCEF]{Policy and Charging Enforcement Function}
\acro{pcrf}[PCRF]{Policy and Charging Rules Function}
\acro{pctl}[PCTL]{Parental Control}
\acro{pep}[PEP]{Performance Enhancement Proxy}
\acro{pgw}[PGW]{Packet Gateway}
\acro{pdn}[PDN]{Packet Domain Network}

\acro{qoe}[QoE]{Quality of Experience}
\acro{qos}[QoS]{Quality of Service}

\acro{reg}[REG]{Regional Network}

\acro{sdn}[SDN]{Software-Defined Networking}
\acro{sfc}[SFC]{Service Function Chaining}
\acro{srv}[SRV]{Server}

\acro{tcp}[TCP]{Transmission Control Protocol}


\acro{vas}[VAS]{Value-Added Service}
\acro{vm}[VM]{Virtual Machine}
\acro{vnf}[VNF]{Virtual Network Function}
\acro{vopt}[VOPT]{Video Optimizer}

\acro{wap}[WAP]{Wireless Application Protocol}
\acro{wapgw}[WAPGW]{Wireless Application Protocol Gateway}
\acro{woc}[WOC]{Web Optimization Controller}
\acro{wopt}[WOPT]{Wide-Area Network Optimizer}




\end{acronym}

\maketitle
\begin{abstract}
While services benefit from distributed cloud centers
running in isolation, allowing multiple centers to cooperate on implementing services 
unlocks the full power of distributed cloud computing. 
Distributed cloud services are typically set up by chaining together a number of functions
that are 
specified with an \emph{implicit} order. They can incorporate complex structures, 
e.g., include 
functions that classify and forward flows over distinct branches and functions that 
are traversed by certain types of flows but skipped by others. 
These requirements need specification techniques more powerful than existing
graph-based ones. We present a context-free grammar for abstract description
of service function chaining structures and a concrete syntax based on
the YANG data modeling language that can easily be translated into an explicit
configuration of service functions. Finally, we present examples of using our
models for complex services within common use cases of service function chaining.
\end{abstract}




\section{Introduction}


According to a cloud-readiness study published in the latest Cisco Global Cloud 
Index~\cite{CiscoGCI2014}, in the coming years, cloud computing operators are expected to
be prepared for a wide range of cloud services including video and music streaming 
in different qualities, communication via text/voice/video, medical and safety 
applications, stock trading, and personal content storage. Some of these services
have very strict requirements that should be considered in all stages of service
design, implementation, deployment, and maintenance. Distributed cloud computing
divides the burden of fulfilling these requirements over multiple cloud centers.
\ac{nfv}, in turn, aims at designing efficient architectures and solutions and
reducing the complexity of service provisioning in cooperating distributed cloud 
centers~\cite{etsi-usecases}.

By applying \ac{nfv} in large-scale networks, a service can be defined as 
a composition of multiple service functions 
that should be traversed by network flows in a specific order~\cite{etsi-terminology}.
The simplest case for such a service is a linear chain of at least one service function 
between two specific endpoints in the network. Inserting functions that can split
network flows over different paths makes the structure of a service 
more complicated than a simple chain. Such services can be modeled as directed 
graphs consisting of service functions as nodes and the connections between pairs of 
service functions as edges of the graph. These graphs are known as \emph{forwarding graphs} 
in ETSI \ac{nfv} terminology \cite{etsi-terminology}.

For resource allocation and network optimization purposes, network operators need
precise and compact representations for the graphs that
model the structure and requirements of a service.  As long as the order of traversing the functions
is fixed and given, traditional graph representations (e.g., adjacency
lists and matrices) can be used for modeling the services. But these representations
can quickly become ineffective when the exact order of traversing the functions
is not specified or not relevant for the functionality of the service. For example, 
when there are no direct dependencies between two functions that should be applied
to a flow, the network can benefit from a flexible service 
representation that allows the operator to compose the required services in the 
most efficient way. We have addressed this problem by introducing a context-free
grammar in a previous work~\cite{mehraghdam2014specifying}. In this paper, 
we present an enhanced version of 
our service description grammar that can express more service structures compared 
to its earlier version. Based on this grammar, we propose an extension to the 
YANG~\cite{rfc6020} data model for service function chaining, published as a 
working draft~\cite{draft-penno-sfc-yang-11} by the IETF \ac{sfc} group. 
With this extension, complex and flexible service compositions can be defined, 
configured, and reused.

This paper is organized as follows: after a brief overview of related work in
Section~\ref{sec:relatedwork}, we describe our models in Section~\ref{sec:specification}.
To highlight the value of our proposed service specification models, we present 
examples of complex chains from common service function chaining use cases in
Section~\ref{sec:usecases}, before we conclude the paper in Section~\ref{sec:conclusion}.

\section{Related Work}
\label{sec:relatedwork}

Sun et al.\,\cite{sun2012survey} have published a survey of cloud computing description
languages. Our model differs 
from the existing models in the sense that we focus on a flexible description
for expressing how the components of a distributed cloud service are \emph{chained and 
composed} to set up a service. In our previous work~\cite{mehraghdam2014specifying},
we have elaborated on the importance and influences of the service structure on 
different metrics concerning the network operators, service providers, and users.
In this paper, we extend our previous description model with more features making
it capable of describing more complex structural requirements. In a similar 
approach to flexibility in service description, Keller et al.\,\cite{Keller2014b} propose modeling 
a cloud application as a generic graph template that can be modified
and adapted during and after deployment.



The YANG~\cite{rfc6020} data modeling language is 
designed to provide a hierarchical model of configuration and runtime data for the 
Network Configuration Protocol (NETCONF). It is used in various network 
management scenarios and supports data instances in different formats, e.g., XML.
Moreover, application of YANG/NETCONF in software-defined networking,
e.g., within the OpenDaylight~\cite{medved2014opendaylight} architecture, makes it 
a good candidate for being adopted in network function virtualization
management and orchestration. For example, Scholler et al.\,\cite{scholler2013resilient} have 
defined a YANG-inspired model for expressing the resilience and scalability
requirements of \acp{vnf}.
The IETF \ac{sfc} working group is also developing a 
YANG model that currently can express \ac{sfc} scenarios with a linear structure.
Based on the latest draft~\cite{draft-penno-sfc-yang-11} of the IETF \ac{sfc} YANG model,
we have defined our model to also include the description of complex service function
chaining structures.


\section{Flexible Service Specification}
\label{sec:specification}

The initial version of our context-free grammar~\cite{mehraghdam2014specifying} can specify a 
flexible description for the way service functions are 
chained together to build a service. In this section, we present an enhanced and 
more powerful version of this grammar and a YANG data model based on the new grammar. 
The grammar and the data model both describe a service as an arrangement of different
\emph{compositions} of service functions. For simplicity, we assume all the 
functions that build up a service are virtualized, and we use \acp{vnf} and service
functions interchangeably in this paper. A composition in the simplest case is a single
\ac{vnf} or an endpoint of a service flow but it can also be complex like a multi-branch
structure.

\subsection{Context-Free Grammar}
\label{subsec:grammar}

We express our context-free grammar in \ac{ebnf} as shown in Figure~\ref{fig:grammar}.
Terminals of the grammar are given in bold font. $\langle\text{vnf}\rangle$
and $\langle\text{endpoint}\rangle$ correspond to the set of available \acp{vnf}
and the set of points in the network where the service flows
start or end.

This grammar can be used for specifying a set of
\emph{totally} ordered \acp{vnf} to be chained together in the given order (simple sequence) and
a set of \emph{partially} ordered \acp{vnf} to be chained together in the most
efficient order according to the optimization objectives in the network (\textbf{best-binding}).
Extending our previous grammar, a set of \acp{vnf} can now be specified to be
chained together in a way that all possible 
permutations of them are traversable (\textbf{all-bindings}), i.e., a full mesh of
paths has to be built among the \acp{vnf}.
A complex branching structure for splitting the flows over different 
branches (\textbf{split}) can also be expressed. This composition consists of:
\begin{itemize}[noitemsep]
 \item a \ac{vnf} that can classify and split the flows over different branches,
 specified as the first $\langle\text{func}\rangle$ in the \textbf{split} composition,
 \item an optional best-binding composition to be traversed before the flows 
 reach the branches \cite{mehraghdam2014specifying},
 \item branches that can consist of a single \ac{vnf} or endpoint, a composition 
 of multiple \acp{vnf}, or can be an empty branch (\textbf{pass}) that can be used for
 skipping a part of the service structure. In case the branches are identical, 
 they need to be specified only once together with the number of required replications. 
\end{itemize}

Existence of partially ordered sets of \acp{vnf} in a service structure turns the
deployment request for the service into a flexible request that can be translated
into the best possible forwarding graph depending on requirements of the service
and available resources in the network.


\subsection{YANG Data Model}
\label{subsec:yang}

\begin{figure}[!t]
{\scriptsize
  \begin{align*}
  \langle\text{start}\rangle & ::= \textbf{service}~\textbf{\{} \langle\text{comp}\rangle (\textbf{,}\langle\text{comp}\rangle)^*\textbf{\}} \\ 
  \langle\text{comp}\rangle & ::= \langle\text{functions}\rangle  ~|~\langle\text{bestbind}\rangle ~|~\langle\text{allbinds}\rangle ~|~\langle\text{splt}\rangle ~|~\langle\text{func}\rangle \\ 
  \langle\text{bestbind}\rangle & ::= \textbf{best-binding}~\textbf{\{} \langle\text{functions}\rangle \textbf{\}} \\ 
  \langle\text{allbinds}\rangle & ::= \textbf{all-bindings}~\textbf{\{} \langle\text{functions}\rangle \textbf{\}} \\ 
  \langle\text{splt}\rangle & ::= \textbf{split}~\textbf{\{} \langle\text{func}\rangle (\textbf{,}\langle\text{bestbind}\rangle)? (\textbf{;} \langle\text{branch}\rangle)^+ \textbf{\}} \\ 
  \langle\text{branch}\rangle & ::= \langle\text{comp}\rangle (\textbf{,}\langle\text{comp}\rangle)^* (\textbf{.}\langle\text{num}\rangle)? ~|~ \textbf{pass} \\ 
  \langle\text{functions}\rangle & ::= \langle\text{func}\rangle (\textbf{,}\langle\text{functions}\rangle)^* \\ 
  \langle\text{func}\rangle & ::= \langle\text{vnf}\rangle ~|~ \langle\text{endpoint}\rangle \\
  \langle\text{num}\rangle & ::= \langle\text{nonzero}\rangle~\langle\text{digit}\rangle^* \\
  \langle\text{nonzero}\rangle & ::= ~\textbf{1}~|~\textbf{2}~|~\textbf{3}~|~\textbf{4}~|~\textbf{5}~|~\textbf{6}~|~\textbf{7}~|~\textbf{8}~|~\textbf{9} \\
  \langle\text{digit}\rangle & ::= ~\textbf{0}~|~\langle\text{nonzero}\rangle \\
  \langle\text{vnf}\rangle & ::= ~\boldsymbol{f_1}\,|\,\boldsymbol{f_2}\,|\,\dotsb\,|\,\boldsymbol{f_{|F|}} \\ 
  \langle\text{endpoint}\rangle & ::= ~\boldsymbol{p_1}\,|\,\boldsymbol{p_2}\,|\,\dotsb\,|\,\boldsymbol{p_{|P|}} \\ 
  \end{align*}
}
\vspace{-18pt}
\caption{Context-free grammar}
\label{fig:grammar}
\end{figure}

\begin{figure}[!t]
{\scriptsize
\begin{verbatim}
module: flexible-service-specification
 +--rw specification
    +--rw starting-component    component-ref
    +--rw service-component* [component-identifier]
       +--rw component-identifier    string
       +--rw compositions* [composition-identifier]
          +--rw composition-identifier    string
          +--rw (composition-type)?
             +--:(sequence)
             |  +--rw sequence-functions*       service-function
             +--:(best-binding)
             |  +--rw best-binding-functions*   service-function
             +--:(all-bindings)
             |  +--rw all-bindings-functions*   service-function
             +--:(split)
             |  +--rw splitter-function         service-function
             |  +--rw optional-best-binding*    service-function
             |  +--rw outgoing-branches* [branch-id]
             |     +--rw branch-id       uint8
             |     +--rw (branch-type)?
             |        +--:(normal-branch)
             |        |  +--rw composition     component-ref
             |        |  +--rw replications?   uint8
             |        +--:(pass)
             |           +--rw string          string
             +--:(function)
             |  +--rw single-function           service-function
             +--:(link-to-composition)
                +--rw composition               component-ref 
\end{verbatim}
}
\caption{YANG data model}
\label{fig:yangmodel}
\end{figure}

Based on the context-free grammar shown in Figure~\ref{fig:grammar}, we 
define a YANG model for services. YANG data models can be used for defining 
configuration and runtime state of different data elements. Services and the network
functions which composed them can be abstracted as reconfigurable data elements with
certain attributes, e.g., required link capacity and computational resources. 
In a dynamic cloud computing environment, where requirements of services and
availability of network resources are changing over time, the \emph{structure} of a service 
can also be considered as a (re)configurable attribute of the service. 
In the IETF \ac{sfc} YANG data model draft~\cite{draft-penno-sfc-yang-11}, no model has been specified for 
expressing complex service structures besides simple sequence of functions.
To fill this gap, we propose extending the IETF SFC model with our definitions.
We present the tree representation of our YANG module (effectively, the grammar)
for flexible specification of complex services in Figure~\ref{fig:yangmodel} and 
omit the detailed definition of the module because of space limitations.

We have created the tree representation using pyang\footnote{\url{https://code.google.com/p/pyang/}}, an open-source tool for validating YANG
modules. Within this tree, each leaf node that represents a data element is specified by a name and a type, e.g., 
\texttt{identifier} of type \texttt{string}. A list of leaf nodes is
specified by \texttt{<name>*} and type of the leaves, e.g., the list \texttt{sequence-functions}
defines a set of \texttt{service-function}s. Lists that consist of non-leaf nodes
are specified by \texttt{<name>*} and a key written as \texttt{[<name>]} that is unique among all 
items of the list, e.g., in the list of \texttt{compositions} each item has a unique
\texttt{composition-identifier}. Description of the list elements follows as children
of the node. 
To express a choice among different options, the node name is written as
\texttt{(<name>)?} and the possible choices are represented as its child nodes 
with the format \texttt{:(<name>)}.
Optional data items are represented as \texttt{<name>?}, e.g., number of \texttt{replications}
can be optionally specified for a branch if outgoing branches from a splitter 
function are identical.

We use \texttt{service-function} for referring to the service function type defined
in the latest IETF draft of a YANG model for \ac{sfc}~\cite{draft-penno-sfc-yang-11}.
This type corresponds to the \acp{vnf} and endpoints in our grammar description.

Consistent with the context-free grammar,
the YANG module also contains all of the composition types described in 
Section~\ref{subsec:grammar}: \texttt{sequence}, 
\texttt{best-binding}, \texttt{all-bindings}, \texttt{split}, and single \texttt{function}.

In our flexible service specification module, the service
specification consists of a list of \texttt{service-components}, which is a list of 
at least one \texttt{composition} of service functions. 
Definition of components is required for defining complex services, e.g., nested compositions
and branching structures. Such structures can only be expressed using path references within the 
module, as in the YANG data modeling language~\cite{rfc6020} recursive structures are 
not allowed. For this purpose, we define a reference
to \texttt{component-identifier} as a new type called \texttt{component-ref} 
and use it for referring to defined components within other compositions. For 
referring to the starting point of the service among all defined components, we 
use the \texttt{starting-component} reference.

With this model, services can be defined from scratch using the available types of functions in the 
network. Additionally, the operator can keep a catalog of pre-composed services that a 
tenant can request as an standalone service or for using it as part of a more 
complex service structure. Existing services can easily be referenced within a 
service structure using the composition type \texttt{link-to-composition}. 

To keep the descriptions compact, we have only included structural attributes in the
YANG model shown here. Different requirements and specifications of services and 
their components can be included in the model, e.g., as described in the
IETF \ac{sfc} YANG data model draft~\cite{draft-penno-sfc-yang-11}.

\section{Usage Scenarios}
\label{sec:usecases}

For delivering a service, different \acp{vnf} need to be deployed 
in the network and corresponding flows need to be routed through them in a specific 
order, resulting in different service compositions (described in Section~\ref{sec:specification}).
In this section, we give an overview of these compositions occurring in common 
cloud services. We show examples from use cases of service function
chaining~\cite{broadbandforum, draft-ietf-sfc-use-case-mobility-03,
draft-ietf-sfc-dc-use-cases-02, draft-liu-sfc-use-cases-08} in fixed and mobile
broadband networks and data center networks, where distributed cloud services 
can be integrated. We show sample forwarding graphs and how the graphs can be compactly
described using our grammar. Every abstract service description expressed by the 
context-free grammar described in Section~\ref{subsec:grammar} can also be expressed 
using corresponding YANG definitions shown in Section~\ref{subsec:yang}. 

\begin{figure}[!t]
\centering
\subfloat[Adapted from Ref.\,\protect\cite{broadbandforum}]{\includegraphics[width=0.58\linewidth]{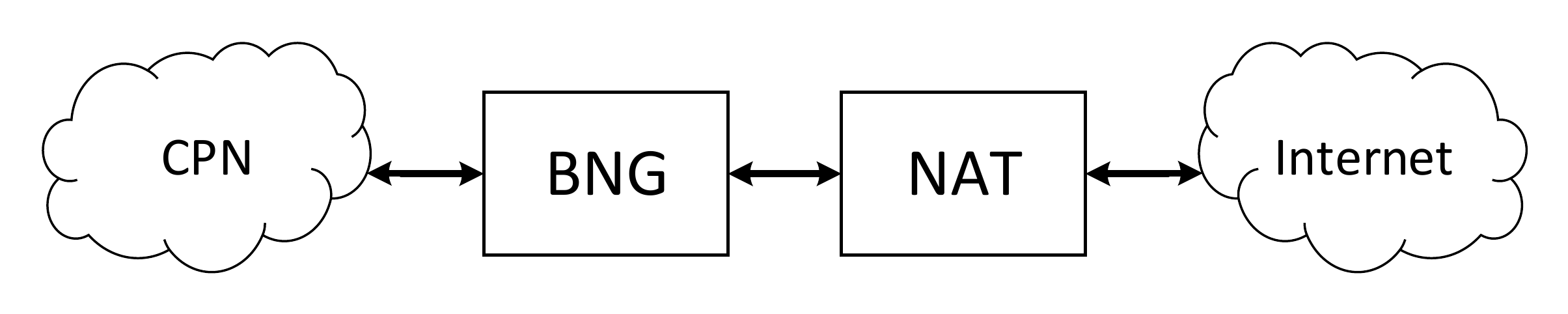}
\label{fig:bng-nat}}
\vspace{-2pt}
\subfloat[Adapted from Ref.\,\protect\cite{broadbandforum}]{\includegraphics[width=0.68\linewidth]{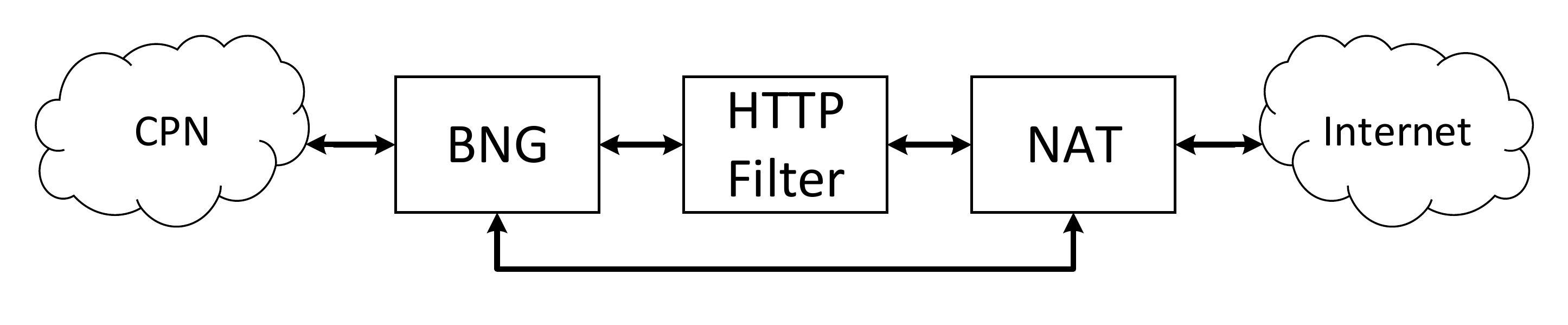}
\label{fig:bng-nat-http}}
\vspace{0pt}
\subfloat[Adapted from Ref.\,\protect\cite{draft-ietf-sfc-use-case-mobility-03}]{\includegraphics[width=0.58\linewidth]{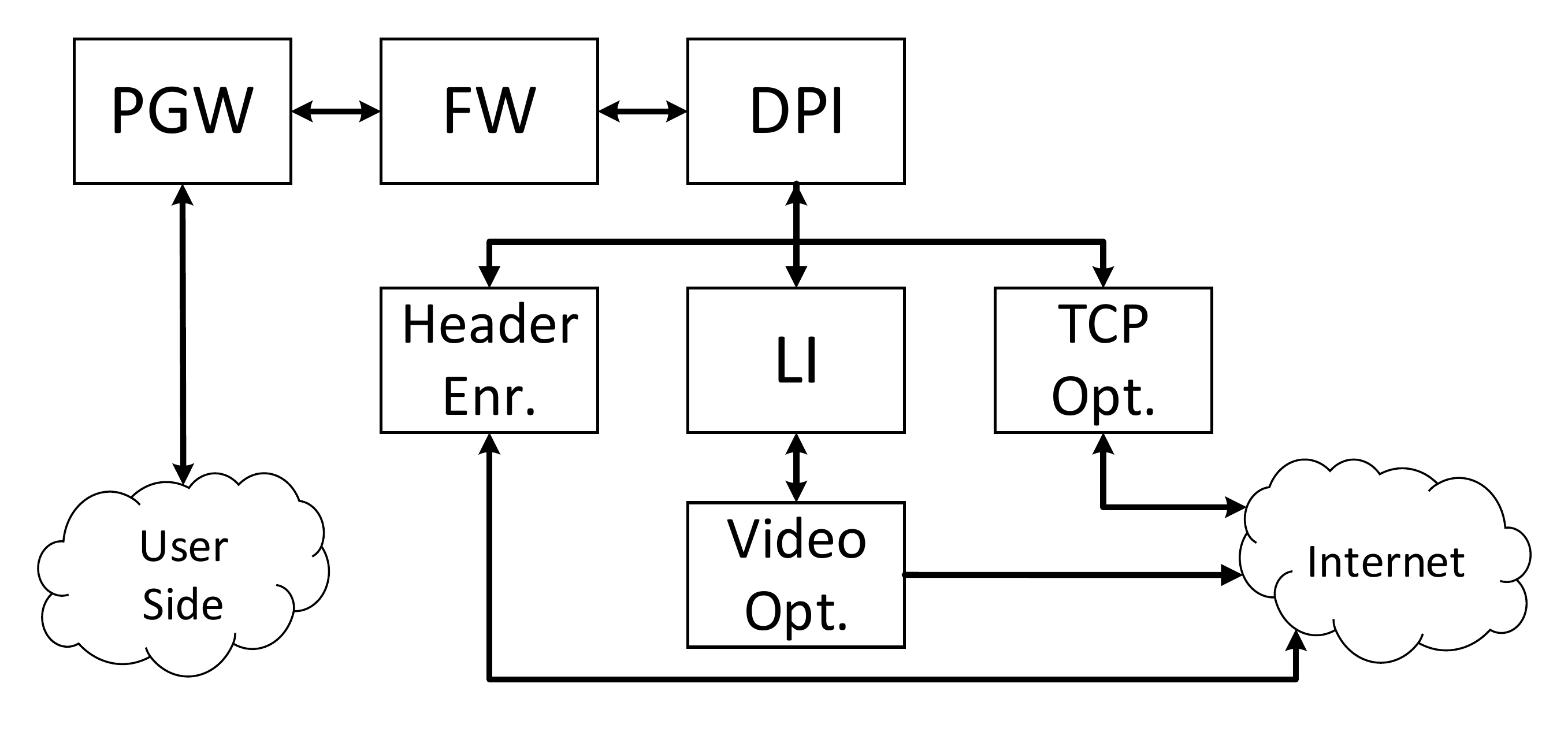}
\label{fig:complex}}
\vspace{-2pt}
\subfloat[Adapted from Ref.\,\protect\cite{draft-ietf-sfc-dc-use-cases-02}]{\includegraphics[width=0.88\linewidth]{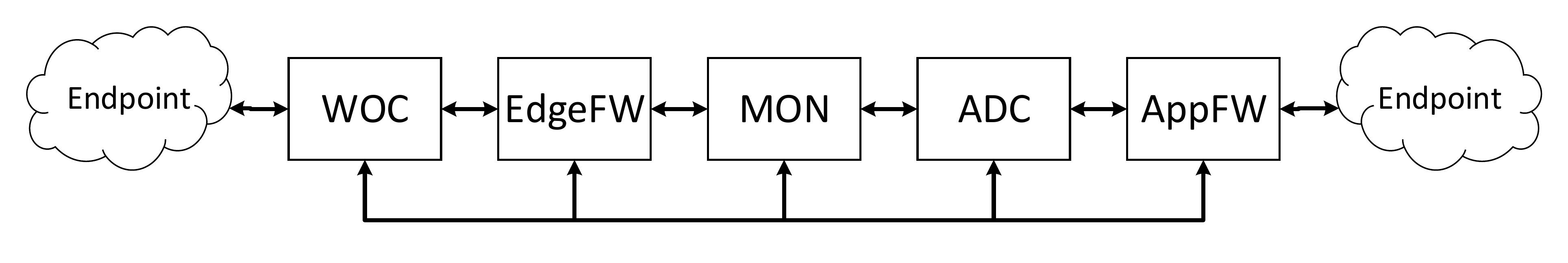}
\label{fig:allbinding}}

\caption{Example service function chaining cases}
\end{figure}

A simple example
of service function chaining in fixed broadband networks consists of a \ac{bng} and a 
\ac{nat} as endpoints of service flows between the customer premises network and public
Internet. Assuming that all network flows need to traverse these two functions, 
Figure~\ref{fig:bng-nat} shows the structure of this service. Using a simple 
sequence from our model, the abstract description of this service is:

{\scriptsize
\vspace{-8pt}
\begin{equation*}
\textbf{service} \textbf{\{} \text{BNG} \textbf{,} ~\text{NAT} \textbf{\}} 
\end{equation*}
}%


Depending on the requirements of the service, this description can be interpreted
as a symmetric service, i.e., the functions are traversed by flows in both directions
between the Internet and the customer network, or it can describe an asymmetric 
service to be applied \emph{only} to flows going from the customer network towards the 
Internet. 

In case the order of traversing the functions does not affect the functionality
of the service, describing the service using a ``best-binding'' composition instead
of a simple sequence of functions allows the operator to order the functions in 
the most beneficial way for the network and the service, e.g.,:

{\scriptsize
\vspace{-8pt}
\begin{equation*}
\textbf{service} \textbf{\{} \textbf{best-binding} \textbf{\{} \text{BNG} \textbf{,} ~\text{NAT} \textbf{\}} \textbf{\}} 
\end{equation*}
}%

Now we assume that HTTP traffic is detected and sent through an HTTP filter function and 
non-HTTP traffic is routed directly between \ac{bng} and \ac{nat}. The corresponding
graph for this service is shown in Figure~\ref{fig:bng-nat-http}. This service 
can be expressed as follows using our ``split'' composition type comprising a branch type
``pass'' to enable skipping the HTTP filter function for some flows:

{\scriptsize
\vspace{-8pt}
\begin{equation*}
\textbf{service} \textbf{\{} \textbf{split} \textbf{\{} \text{BNG} \textbf{;} ~\text{HTTP-Filter} \textbf{;} ~\textbf{pass} \textbf{\}} \textbf{,} ~\text{NAT} \textbf{\}} 
\end{equation*}
}%

The first function in the ``split'' composition (BNG) is the splitter
function and the HTTP-Filter and the ``pass'' keyword each correspond to one outgoing branch.



As a more complex example, we look at a scenario in a mobile broadband network in
Figure~\ref{fig:complex}.
This is a symmetric service between the Internet and a \ac{pgw} where the user
equipments are connected via the access network. \ac{fw} and \ac{dpi} are
applied to all flows and later on the flows are divided over three branches. 
TCP flows need to traverse a TCP optimizer function, flows belonging to a 
certain video streaming service go through a \ac{li} and a video optimizer function,
and other flows need to go through a header enrichment function. This structure 
can be expressed as follows, using a ``split'' composition:

{\scriptsize
\vspace{-8pt}
\begin{align*}
& \textbf{service} \textbf{\{} \\ 
& \quad \text{PGW} \textbf{,} ~\text{FW} \textbf{,} \\
& \quad\textbf{split} \textbf{\{} \text{DPI} \textbf{;} ~\text{Header-Enr} \textbf{;} ~\text{LI} \textbf{,} ~\text{Video-Opt} \textbf{;} ~\text{TCP-Opt} \textbf{\}} \\
& \textbf{\}}
\end{align*}
}%


As a last example, we look at a service chaining scenario in a data center network,
shown in Figure~\ref{fig:allbinding}. Different flows need to traverse different
subsets and different permutations of the set of functions, including a \ac{woc},
a firewall responsible for external threats (EdgeFW), a network and application 
monitoring function (MON), an \ac{adc}, and an application-specific firewall
(AppFW). This complex structure can be compactly described using an ``all-bindings''
composition:

{\scriptsize
\vspace{-8pt}
\begin{equation*}
\textbf{service} \textbf{\{} \textbf{all-bindings} \textbf{\{} \text{WOC} \textbf{,} ~\text{EdgeFW} \textbf{,} ~\text{MON} \textbf{,} ~\text{ADC} \textbf{,} ~\text{AppFW} \textbf{\}} \textbf{\}} 
\end{equation*}
}%

These examples confirm the existence of complex chaining structures in
networks. Our models facilitate the flexible and compact description of possible 
chaining scenarios.

\section{Conclusion}
\label{sec:conclusion}
The grammar presented in this paper provides a powerful tool for abstract 
specification of complex service structures that can be used as a basis for defining 
specification languages and models. The YANG model we have defined based on this 
grammar extends the structural aspects of existing YANG definitions for service 
function chaining that are under development by standardization bodies. These 
definitions and various use case documents prove the existence of service structures
that are more complicated than a linear and totally ordered set of functions, which
makes our model an important contribution towards flexible specification
of complex structures in distributed cloud services. 

\section{Acknowledgments}
This work is partially supported by the German Research Foundation (DFG) within
the Collaborative Research Center “On-The-Fly Computing” (SFB 901) and the 
International Graduate School ``Dynamic Intelligent Systems''.
\\
\\
%
\bibliographystyle{abbrv}
\bibliography{ref}  
%
%

\end{document}